\documentclass[preprint,showpacs,byrevtex,prl,10pt,twocolumn]{revtex4}%
\usepackage{amsfonts}
\usepackage{amsmath}
\usepackage{amssymb}
\usepackage[dvips]{graphicx}%
\providecommand{\U}[1]{\protect\rule{.1in}{.1in}}
\newtheorem{theorem}{Theorem}
\newtheorem{acknowledgement}[theorem]{Acknowledgement}

\begin{document}
\preprint{ }
\title{The Electron Glass in a three-dimensional system}
\author{Z. Ovadyahu}
\affiliation{Racah Institute of Physics, The Hebrew University, Jerusalem 91904, Israel }


\begin{abstract}
We report on non-equilibrium transport features observed in experiments using
three-dimensional amorphous indium-oxide films. It is demonstrated that all
the features that characterize intrinsic electron-glasses which heretofore
were seen in two-dimensional samples are also observed in field-effect
measurements of systems that exhibit three-dimensional variable-range-hopping.
In particular, a memory-dip is observed in samples configured with gate. The
memory-dip width and magnitude support models that associate the phenomenon
with the Coulomb-gap. The memory-dip and the glassy effects disappear once the
quenched disorder in the system is reduced and the system becomes diffusive.
This happens when the Ioffe-Regel dimensional parameter k$_{\text{F}}\ell$
exceeds $\approx$0.3 which is the critical value for the metal-to-insulator
transition in all versions of the amorphous indium-oxides [Phys. Rev. B
\textbf{86}, 165101 (2012)]. This confirms that being in the Anderson
localized phase is a pre-requisite for observing the memory-dip and the
associated glassy effects. The results of the gating experiments suggest that
the out-of equilibrium effect caused by inserted charge extend over spatial
scales considerably larger than the screening length.

\end{abstract}
\maketitle

\section{Introduction}

The out-of-equilibrium transport properties of several Anderson insulators
exhibit glassy features such as slow relaxation, slow approach to a steady
state, ageing, and other memory effects. These non-equilibrium features were
theoretically anticipated and are believed to result from the interplay
between disorder and interactions \cite{1,2,3}. This glassy phase, commonly
referred to as electron-glass, arises from the competition between quench
disorder and Coulomb interactions. It is a generic phenomenon that ought to
apply to all degenerate Fermi systems with localized states interacting via a
Coulomb potential. Experimental evidence for glassy effects, with slow
relaxation of many seconds, has been somewhat scarce and was only reported in
two-dimensional (2D) systems \cite{4}, with thickness d$\approx$180-200\AA .

In this work we extend the study to thick amorphous indium-oxide
(In$_{\text{x}}$O) films that, for sufficiently strong disorder, exhibit 3D
hopping conductivity at liquid helium temperatures. The out-of equilibrium
properties of these samples exhibit glassy effects characteristic of
previously studied electron-glasses These include logarithmic relaxation of
the conductance following excitation by a variety of means, and a memory-dip
(MD) that has all the earmarks of intrinsic electron-glasses.

The memory-dip disappears at the metal-insulator transition of the material,
consistent with the conjecture that the glassy phase is a property peculiar to
the Anderson insulating phase.

Of special focus in this work is the question of how the insertion of charge
into an interacting electronic system takes it out of equilibrium, and in
particular how far the disturbance created in the process propagates into the
system. We describe the results of a special protocol involving a combination
of excitation agents to deal with these questions. It is shown the
perturbation due to charge-insertion propagates deeper into the system than
the screening length. The implications of these results to the slow dynamics
of the electron-glass are discussed.

\section{Experimental}

\subsection{Samples preparation and characterization}

Several batches of amorphous indium-oxide (In$_{x}$O) films were prepared in
this study. These were e-gun evaporated onto room-temperature substrates using
99.999\% pure In$_{2}$O$_{3}$ sputtering target pieces. Two types of
substrates were used; 1mm-thick microscope glass-slides, and 0.5$\mu$m
SiO$_{\text{2}}$ layer thermally grown on
$<$%
100%
$>$
silicon wafers. The Si wafers were boron-doped (and exhibited P-type
conductivity) with bulk resistivity $\rho\simeq$ 2$\cdot$10$^{\text{-3}}%
\Omega$cm, deep into the degenerate regime. These wafers were used as the gate
electrode in the field-effect experiments. The microscope glass-slides were
mostly used for optical characterization and for Hall-Effect measurements.

Deposition was carried out at the ambience of (1-3)$\cdot$10$^{\text{-4}}$
Torr oxygen pressure maintained by leaking 99.9\% pure O$_{\text{2}}$ through
a needle valve into the vacuum chamber (base pressure $\simeq$10$^{\text{-6}}$
Torr). Rates of deposition used for the samples reported here were typically
0.6-0.9~\AA /s. Under these conditions, the In$_{\text{x}}$O samples had
carrier-concentration n in the range (7-8)$\cdot$10$^{\text{19}}%
$cm$^{\text{\={-}3}}$ as measured by Hall-Effect at room temperatures. Film
thickness in this study was 750$\pm$20\AA . Rate of deposition and thickness
were measured by a quartz thickness monitor calibrated using optical
interference measurements on thick MgF$_{\text{2}}$ films.

As-deposited samples had room-temperature resistivity $\rho$ in excess of
10$^{\text{5}}\Omega$cm which, for the low temperature studies, had to be
reduced by several orders of magnitude. This was achieved by thermal
annealing. A comprehensive description of the annealing process and the
associated changes in the material microstructure are described elsewhere
\cite{5,6,7}. For completeness, we give here a brief summary of the changes
occurring during the annealing process.

The main effect of annealing is a $\simeq$10\% increase of material density.
The thickness shrinkage upon annealing was directly measured by x-ray
interferometry and indirectly as an increase of refractive index \cite{7}. In
addition, the visibility of the x-ray interference was highly enhanced
following extended annealing, presumably due to the smoothing-out
potential-fluctuations \cite{6}. The main change observed in the optical
properties of the material is a decrease of the optical-gap \cite{7}. This
apparently reflects the enhanced inter-atomic overlap associated with the
density increase, leading in turn to wider bands.

The annealing protocol that was used in the current study was as follows:
After deposition and initial conductance measurement that, in the initial
stages, required the use of electrometer (Keithley 617), the sample was
attached to a hot-stage at a constant temperature T$_{\text{a}}$, typically
5-10 degrees above room temperature. The resistance R of the sample was
observed to slowly decrease over time. T$_{a}$ was raised by few degrees
whenever $\Delta\rho/\rho$ over 24 hours was less than 1\% (and the value of
the resistance was still higher than desired). To obtain a sample with $\rho$
that was useful for the measurements reported here took 10-20 thermal cycles.
The annealing temperature T$_{a}$ was limited to $\approx$360K to minimize the
risk of crystallization. Results of measurement on the series of samples
reported in this study were generated from two batches of deposition. The
samples used for the low temperature measurements were made from deposited 1mm
wide strips. These were cut into $\approx$2mm long pieces and indium-contacts
were pressed at their ends for electrical connections. Sample length L was
typically 1mm.

Hall Effect measurement were carried out on samples that were patterned in a
6-probe configuration using stainless-steel masks. These were prepared during
the same deposition as the strips used for the low temperature transport
measurements. A standard Hall-bar geometry was used with the active channel
being a strip of 1 mm wide, and 10 mm long. The two pairs of voltage probes
(that doubled as Hall-probes), were spaced 3 mm from one another along the
strip. This arrangement allowed us to assess the large scale uniformity of the
samples, both in terms of the longitudinal conductance and the Hall effect.
Excellent uniformity was found on these scales; resistivities of samples
separated by 1~mm along the strip were identical to within $\pm$5\%. No change
(within the experimental error of 3\%) was observed in the hall effect due to
annealing (tested for samples with room temperature resistivity smaller than
$\simeq$0.4$\Omega$cm which was the highest $\rho$ in the samples studied in
this work).

\subsection{Measurements techniques}

Conductivity of the samples was measured using a two-terminal ac technique
employing a 1211-ITHACO current pre-amplifier and a PAR-124A lock-in
amplifier. Except when otherwise noted, measurements reported below were
performed with the samples immersed in liquid helium at T=4.1K maintained by a
100 liters storage-dewar, which allowed long term measurements of samples as
well as a convenient way to maintain a stable temperature bath. The ac voltage
bias was small enough to ensure linear response conditions (judged by Ohm's
law being obeyed within the experimental error).

As in a previous study \cite{6}, we use in this work as a dimensionless
measure of disorder: k$_{\text{F}}\ell$=(9$\pi^{\text{4}}/$n)$^{\text{1/3}%
}\frac{\text{R}_{\text{Q}}}{\rho_{\text{RT}}}$ (R$_{\text{Q}}$=$\hslash
/$e$^{\text{2}}$). This is based on free-electron expressions using the
measured room-temperature resistivity $\rho_{\text{RT}}$ and the
carrier-concentration n, obtained from the Hall-Effect measurements, as
parameters. More details of preparation and characterization of In$_{\text{x}%
}$O samples are given elsewhere \cite{5,6,7}.

Optical excitation was accomplished by exposing the sample to AlGaAs diode
(operating at $\approx$0.88$\pm$0.05$\mu$m), placed $\approx$15mm from the
sample. The diode was energized by a computer-controlled Keithley 220
current-source. The samples were attached to a probe equipped with calibrated
Ge and Pt thermometers and were wired by triply-shielded cables to BNC
connectors at room temperatures. The effective capacitance of the wires was
$\leq$20pF but the sample-gate capacitance in the MOSFET-like samples used for
the field effect measurements was an order of magnitude larger. This still
allowed the use of 40-75Hz ac technique except when the sample resistance
exceeded 20M$\Omega$. In the latter case a frequency of 4.8Hz had to be used
with some compromise on the signal-noise ratio.

Fuller details of measurement techniques are given elsewhere \cite{8}.

\section{Results and discussion}

\subsection{Hopping conductivity and current-voltage characteristics}

The behavior of the near-equilibrium (conductance) and steady-state
(current-voltage characteristics) properties of the 750\AA \ indium-oxide
films used in this work was typical of a variable-range-hopping (VRH)
mechanism. Conductance versus temperature G plots of two samples are shown in Fig.1.%

\begin{figure}[ptb]%
\centering
\includegraphics[
trim=0.000000in 0.000000in -0.015536in 0.000000in,
height=2.5832in,
width=3.4255in
]%
{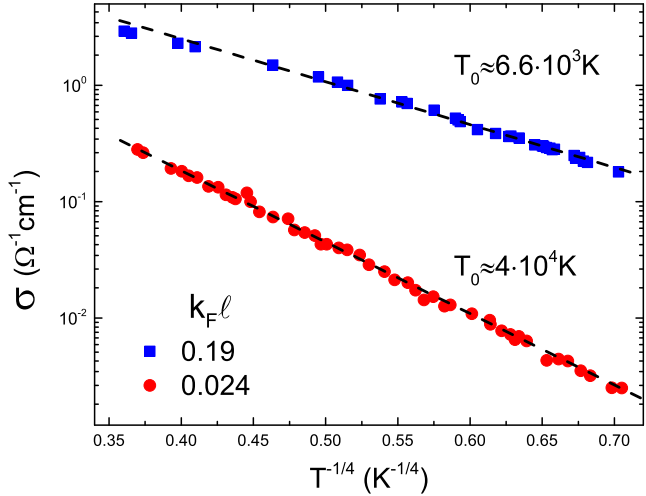}%
\caption{(color online) Conductivity as function of temperature plots for two
of the studied samples (labeled by their k$_{\text{F}}\ell$ values). The
curves are marked by the characteristic temperature T$_{\text{0}}$ based on
the respective logarithmic slope.}%
\end{figure}

At the 4-35K temperature range G(T) of these samples is consistent with a
Mott-type VRH for a three-dimensional (3D) system \cite{9}. Using the
theoretical expressions \cite{9,10} for G(T)$\propto\exp$[-(T$_{\text{0}}%
$/T)$^{\text{1/4}}$], the associated localization length $\xi$ may be obtained
from \cite{8} k$_{\text{B}}$T$_{\text{0}}\cong$3$\cdot$($\partial$%
n$/\partial\mu\cdot\xi^{\text{3}}$)$^{\text{-1}}$ where k$_{\text{B}}$ is the
Boltzmann constant and $\partial$n$/\partial\mu$ is the material's
thermodynamic density of states (DOS) at the chemical potential. For
indium-oxide with carrier-concentration n$\approx$8$\cdot$10$^{\text{19}}%
$cm$^{\text{-3}}$ $\partial$n$/\partial\mu\approx$2$\cdot$10$^{\text{32}}%
$erg$\cdot$cm$^{\text{-3}}$ which then gives $\xi\simeq$14\AA \ and $\xi
\simeq$26\AA \ for the samples depicted in Fig.1. The hopping length r(T)
$\simeq\xi{}$[(T$_{\text{0}}$/T)$^{\text{1/4}}$] for these samples is
$\approx$140\AA \ and $\approx$170\AA \ respectively at the lowest temperature
studied here (T=4.1K), consistent with 3D variable-range-hopping. It should be
remarked that these VRH expressions are based on simplified models \cite{11}
and therefore our estimates for $\xi$ and r may be off by a certain factor.
Note however, that our estimate of $\xi$ for the more disordered sample in
Fig.1 is close to the Bohr radius for indium-oxide ($\simeq$10\AA ), which is
what one expects the localization-length to be for a deeply insulating sample,
so our estimates of $\xi$ are probably not so bad.

Another characteristic manifestation of hopping conductivity is the
sensitivity of G to an applied electric field. In Fig.2 such a behavior is
shown for 3 samples having different degrees of disorder (labeled by their
k$_{\text{F}}\ell$ value).%

\begin{figure}[ptb]%
\centering
\includegraphics[
height=2.292in,
width=3.4255in
]%
{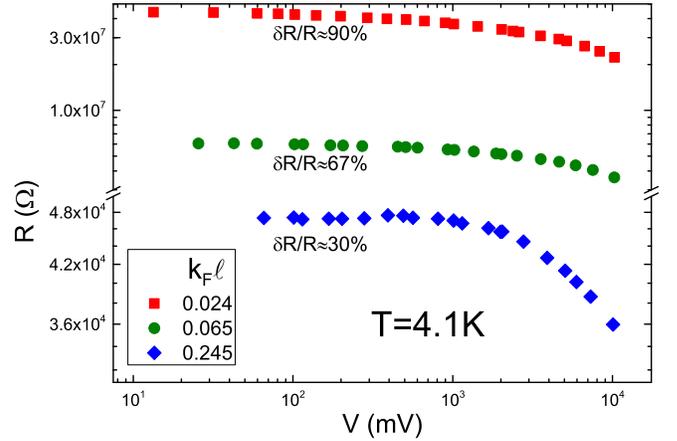}%
\caption{(color online) The dependence of the resistance on the applied
voltage for three typical samples illustrating the non-Ohmic behavior. The
length of all samples L=1$\pm$0.1mm. The curves are labeled by the relative
change of the resistance, $\delta$R/R$\equiv$[R$_{\text{0}}$%
-R(10V)]/R$_{\text{0}}$ where R$_{\text{0}}$ is the resistance measured at the
Ohmic voltage.}%
\end{figure}

The data in this figure illustrates the dependence of G on voltage. Theories
anticipate \cite{12,13,14,15} deviations from linear-response when the applied
field F exceeds (k$_{\text{B}}$T)/eL where L is of the order of \cite{14} r or
the percolation radius \cite{12} L$_{\text{C}}$. It is commonly found however
that non-ohmicity sets-in at a considerably smaller F than can be attributed
to any of these lengths \cite{16,17}. The current case is no exception;
deviation from Ohm's law are observable in these data in fields as small as
10V$\cdot$m$^{\text{-1}}$ which imply length scales of several microns. Note
also that the higher is the disorder (and thus the smaller are both r and
L$_{\text{C}}$), the smaller is the voltage required to maintain Ohmic
behavior. Based on the sample-size dependence for non-Ohmicity in a number of
systems, it has been suggested that the anomalous sensitivity of disordered
systems to applied fields result from the presence of long range potential
fluctuations, which is also revealed in the dependence of non-Ohmic effects on
sample-size \cite{18}.

\subsection{Out of equilibrium transport properties}

The first indication for non-ergodic behavior is the temporal evolution of G
following a sudden cooling process, which is actually the initial stage of
every new experiment in this field. All intrinsic electron-glasses exhibit
G(t)$\propto$G$_{\text{0}}$- a$\cdot\log$(t) law after a quench-cooling from
room-temperature to a cold enough bath. This usually is carried out in stages;
first, the sample, attached to a probe, is cooled to 80-100K and held there
for few minutes to lower the heat capacity of the sample-stage and surrounding
devices (thermometers, infrared-LED, etc.). Then, the probe is quickly lowered
to the liquid helium bath. Measurement of the conductance relaxation versus
time may commence once the attached thermometer confirms that a stable
temperature has been reached. An example of a typical run is shown in Fig.3.%

\begin{figure}[ptb]%
\centering
\includegraphics[
height=2.5201in,
width=3.4247in
]%
{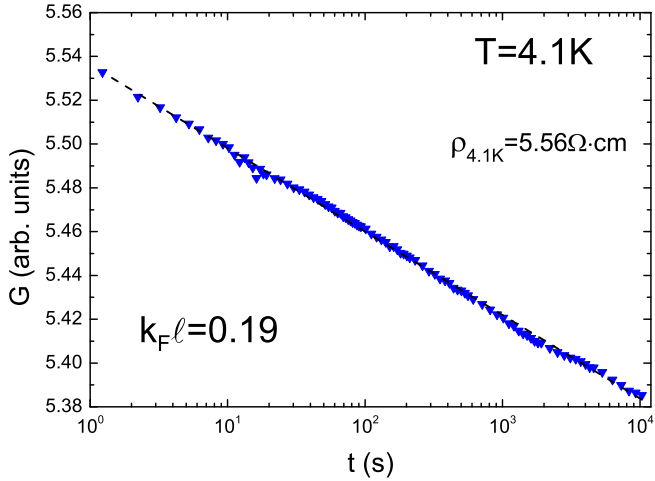}%
\caption{(color online) Temporal dependence of conductance relaxation
following quench-cooling from high temperature to the liquid helium bath. }%
\end{figure}

The prefactor a in G(t) is related to the magnitude of the nonequilibrium
effects that may be expected for a given sample (relative to its `equilibrium'
conductance). As will be discussed below, this information may be useful when
evaluating some aspects of the mechanism by which gating the system (or
radiating it with infrared source) takes it far from the equilibrium.

Of the many ways to take the system out of the equilibrium the field-effect is
arguably the least "destructive" and the most widely used technique. The
field-effect also exposes the identifying signature of the electron-glass - a
memory-dip (MD)\ which appears a minimum in the conductance versus gate
voltage G(V$_{\text{g}}$) at the gate voltage where the system was allowed to
relax. It has been conjectured by several authors \cite{19,20,21} that the MD
is a reflection of the Coulomb gap \cite{22,23,24}; inserting (removing)
particles to (from) the system randomizes the electronic system thus
destroying the Coulomb-gap (if randomization is complete) and enhancing the
conductance. Indeed many of the experimentally observed features seem to be
accounted for by this approach \cite{4}. There are questions however as to the
underlying excitation mechanism, in particular, it is interesting to know what
is the spatial range of the disturbance introduced by a sudden change of the
gate voltage. In the next few paragraphs these issues will be discussed using
data for the infrared excitation and field effect measurements.

The field effect of one of the studied samples is shown in Fig.4 at different
temperatures.%
\begin{figure}[ptb]%
\centering
\includegraphics[
height=2.4259in,
width=3.4255in
]%
{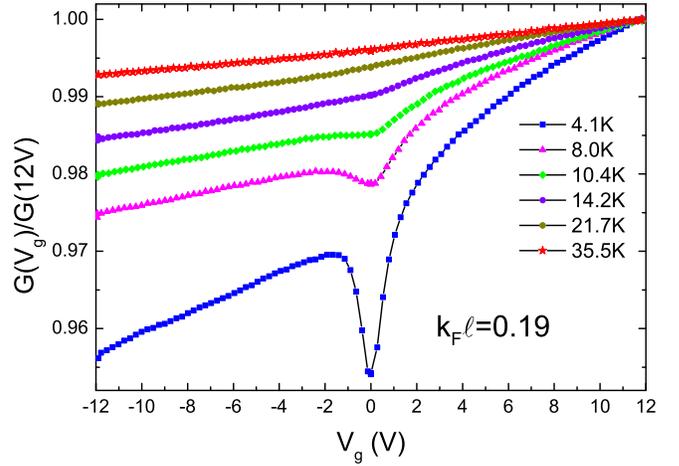}%
\caption{(color online) The dependence of the field effect characteristics on
the bath temperature for the same sample as in Fig.1 and Fig.3. Note that the
magnitude of the MD decreases very fast with temperature relative to the
change in the anti-symmetric component.}%
\end{figure}

These data, taken on one of the samples depicted in Fig.1, illustrates the
evolution of the field effect traces G(V$_{\text{g}}$) with temperature. Each
curve is composed of an asymmetric (with respect to the equilibrium gate
voltage) G(V$_{\text{g}}$), reflecting the thermodynamic density of states,
and a MD that quickly diminishes in magnitude with temperature. The overall
behavior is qualitatively similar to the previously studied temperature
dependence of a 2D systems \cite{8,25}. However, the larger range of the
V$_{\text{g}}$ scan here reveals that the modulation produced by the MD
extends much further than the `half-height' width consistent with a point made
earlier \cite{26}. It also suggests that the thermodynamic part of
G(V$_{\text{g}}$) is a somewhat concave curve presumably due to a
disorder-modified $\partial$n$/\partial\mu\{\varepsilon\}$ at the tail of the
conduction-band. The Fermi energy E$_{\text{F}}$ associated with n$\simeq
$8$\cdot$10$^{\text{19}}$cm$^{\text{-3}}$ is near the bottom of the conduction
band as illustrated in the schematic drawing in Fig.5. The figure shows
G(V$_{\text{g}}$) plots for two samples measured under conditions where the MD
contribution is negligible and it is interesting to note that the average
slope is stronger for the more disordered sample (even though it was measured
at a higher temperature).%
\begin{figure}[ptb]%
\centering
\includegraphics[
height=2.3993in,
width=3.4255in
]%
{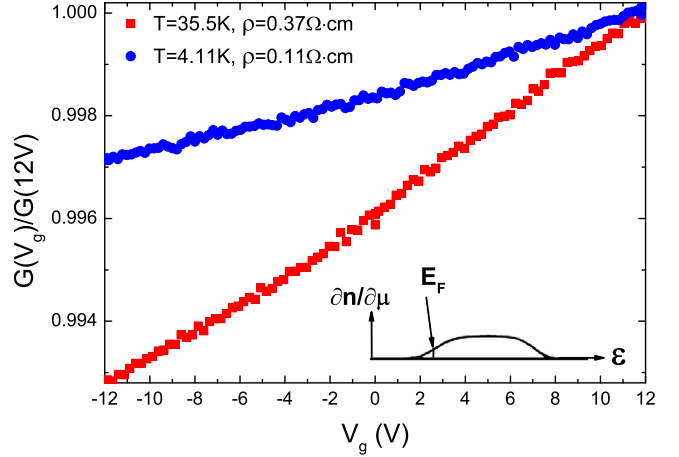}%
\caption{(color online) The field effect for two samples measured under the
conditions that the MD is unresolved (within the experimental error). Traces
are labeled by the measurement temperature and the sample resistivity at this
temperature. Squares are for samples with k$_{\text{F}}\ell$=0.19 while
circles are for sample with k$_{\text{F}}\ell$=0.28. The schematics in the
graph illustrates the origin of the concave form of the thermodynamic
component of the field-effect G(V$_{\text{g}}$). }%
\end{figure}

We have used such a non-linear extrapolated curve as a baseline in estimating
the relative magnitude $\Delta$G/G of the MD for a given G(V$_{\text{g}}$) as
illustrated for example in Fig.6 for one of the samples. The inset to the
figure depicts the dependence of the MD magnitude on the sample resistivity
(all measured with the same sweep rate at the bath temperature of 4.1K and
24-30 hours after their cool-down).
\begin{figure}[ptb]%
\centering
\includegraphics[
height=2.2459in,
width=3.339in
]%
{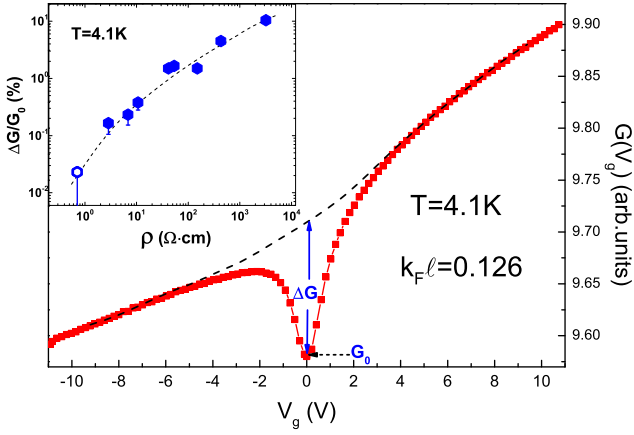}%
\caption{(color online) The field effect for one of the samples studied. The
conductance components $\Delta$G and G$_{\text{0}}$ that are used in assigning
a relative magnitude for the MD are marked by arrows. The inset shows the
dependence of $\Delta$G/G$_{\text{0}}$ on the sample resistivity at T=4.1K.}%
\end{figure}

The data in the inset show that the MD magnitude vanishes, faster than a
power-law, as $\rho$ decreases. Note that the least resistive sample in this
series is assigned with an empty symbol and a downward-going error-mark to
indicate that its very small MD has a borderline value in terms of our
signal-noise ratio. This sample is actually near the transition to the
metallic phase; its k$_{\text{F}}\ell$ is 0.28 which is within the critical
value for this system (k$_{\text{F}}\ell$)$_{\text{C}}$=0.3$\pm$0.2. Indeed,
as illustrated in Fig.7, $\Delta$G/G is an exponential function of
k$_{\text{F}}\ell$ suggesting that the MD (and the associated non-ergodic
effects) are peculiar to the \textit{insulating} phase and vanish at the
metal-insulator transition where the system crosses over to the diffusive regime.

This is one of the main results of the work.

The typical shape and width of the MD in these 3D samples turn out to be
remarkably similar to these of a 2D samples of amorphous indium-oxide with
comparable carrier-concentration. A comparison between the field effect of two
and three-dimensional samples is given in Fig.8.%

\begin{figure}[ptb]%
\centering
\includegraphics[
height=2.566in,
width=3.4255in
]%
{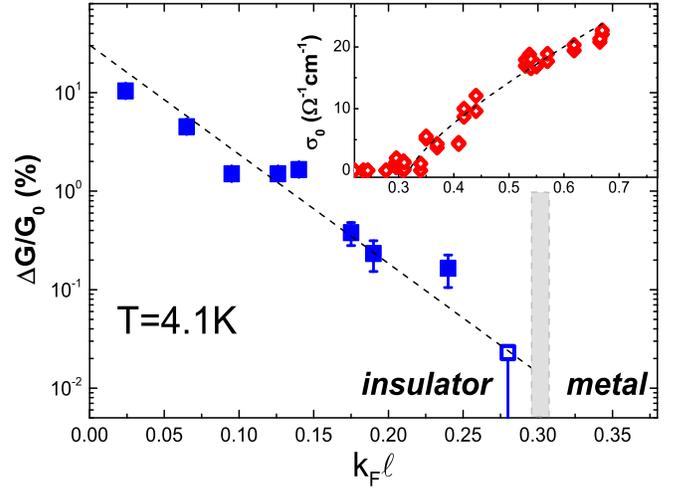}%
\caption{(color online) The relative magnitude of $\Delta$G/G$_{\text{0}}$
using the same data as in the inset to Fig.6 but plotted as function of the
dimensionless parameter k$_{\text{F}}\ell$. The gray rectangle marks the
boundaries of the metal-insulator transition (MIT) of amorphous indium-oxides.
The inset shows the zero temperature conductivity of In$_{\text{x}}$O as
function of k$_{\text{F}}\ell$ (taken from reference \cite{6}). }%
\end{figure}
The nearly identical MD width $\Gamma$ of the thin versus thick samples is
consistent with that the screening-length $\lambda$ taken as ($\pi
e^{\text{2}}\partial$n$/\partial\mu)^{\text{-1/2}}$ is the relevant scale for
assigning an energy scale $\Gamma^{\ast}$ to the $\delta$V$_{\text{g}}$
associated with the MD. In this procedure, used in \cite{4,27}, $\Gamma^{\ast
}$=($\delta$V$_{\text{g}}$)C/[e$\lambda$($\partial$n$/\partial\mu$%
)$_{\text{E}_{\text{F}}}$] where C is the sample-to-gate capacitance (per unit
area), e is the electron-charge, ($\partial$n$/\partial\mu$)$_{\text{E}%
_{\text{F}}}$ is the material DOS at the Fermi energy, and $\delta
$V$_{\text{g}}$ is the range in the G(V$_{\text{g}}$) plot modulated by the
interaction underlying the MD (presumably, the Coulomb interaction). Note that
$\Gamma^{\ast}$ is the shift of the chemical potential (relative to the
band-edge) associated with $\delta$V$_{\text{g}}$.%

\begin{figure}[ptb]%
\centering
\includegraphics[
height=2.3436in,
width=3.4255in
]%
{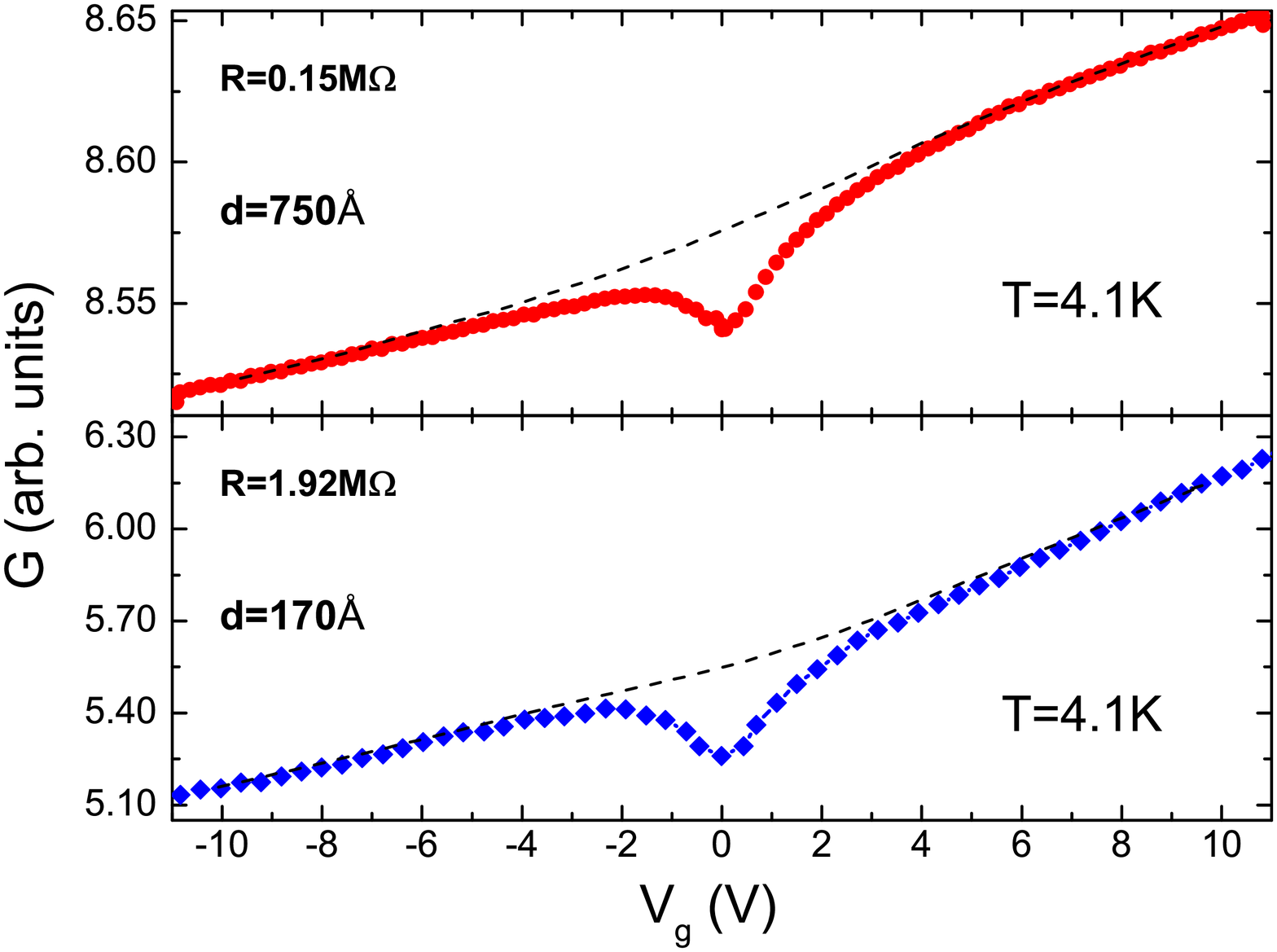}%
\caption{(color online) Comparing the field-effect G(V$_{\text{g}}$) of one of
the 3D samples in this work with a 2D sample of the same material (A. Vaknin,
Ph.D. thesis). Note the similar width of the MD. The carrier concentration n
in the 2D sample was 6.1$\cdot$10$^{\text{19}}$cm$^{\text{-3}}$ and n=8$\cdot
$10$^{\text{19}}$cm$^{\text{-3}}$ for the 3D sample.}%
\end{figure}

We digress now to describe another way to drive the system out of the
equilibrium; excitation by an electromagnetic field, specifically using
infrared radiation. Exciting the system by exposure to infrared radiation will
serve as a tool in a more elaborate gating experiments described below.

The IR-protocol has been used \cite{28} on several electron-glasses with
different degree of efficiency in terms of randomizing the electronic
space-energy configuration (or, simply stated, destroying the Coulomb gap). A
plausible route by which IR exposure destroys the Coulomb gap is a two-step
process \cite{28}: First, electrons are excited from a localized site to a
spatially-overlapping excited state (most likely, an extended state above the
mobility edge). Then, a multi-phonon non-radiative process relaxes the system
involving sufficiently high energy to overcome the Coulomb gap. Such events
should efficiently randomize the electronic configuration. Naturally, there is
also a heating effect accompanying the IR radiation, including indirect
heating through the sample immediate environment (substrate, sample-stage, the
helium bath being heated by the IR LED-case, etc.). It was however
demonstrated experimentally that the MD may be completely washed-out by a
brief exposure to IR while heating could be ruled out as a significant factor
\cite{29}. The heating effect may be separated from the optical excitation
part: Moderate heating (raising the bath temperature by $\Delta$T/T%
$<$%
1), induces a slow process requiring a long time for the randomization to
become appreciable \cite{30}. This is so because only the high energy phonons
in the Bose-Einstein distribution tail are relevant for the process. The
degree of randomization may be tested by performing a quick G(V$_{\text{g}}$)
sweep; absence of the MD indicates complete randomization.

The results of two IR protocols, one where the IR exposure is brief, and the
other where it is allowed to operate on the system for an extended time are
shown in Fig.9 and Fig.10.%

\begin{figure}[ptb]%
\centering
\includegraphics[
height=2.5028in,
width=3.4255in
]%
{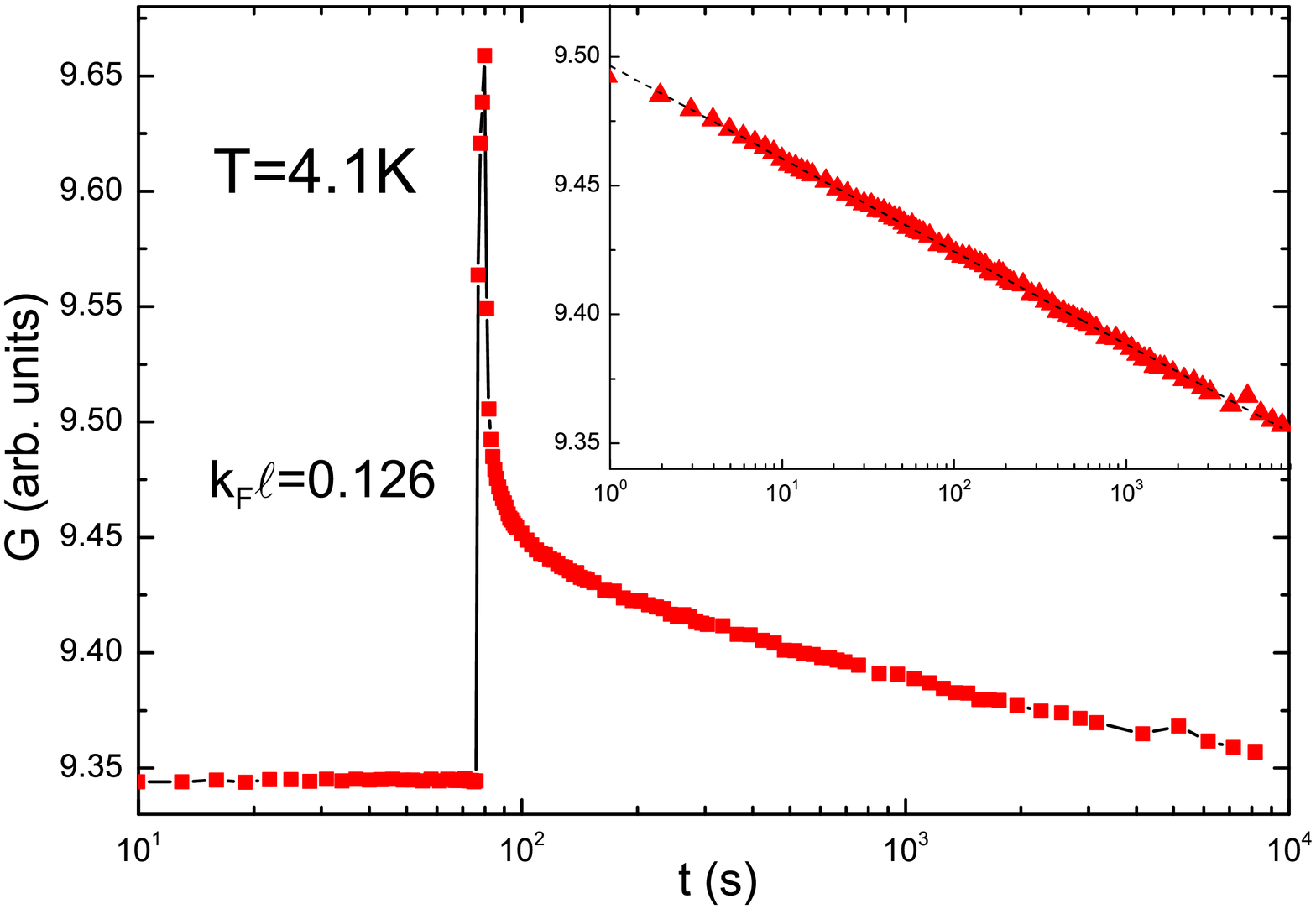}%
\caption{(color online) The temporal evolution of the conductance G(t) during
the IR protocol. The 50mW infrared source is turned on for 2 seconds after
$\simeq$80s baseline trace is recorded. After the IR sources is turned off,
G(t) is observed to be logarithmic as illustrated in the inset.}%
\end{figure}
%

\begin{figure}[ptb]%
\centering
\includegraphics[
height=2.7532in,
width=3.4255in
]%
{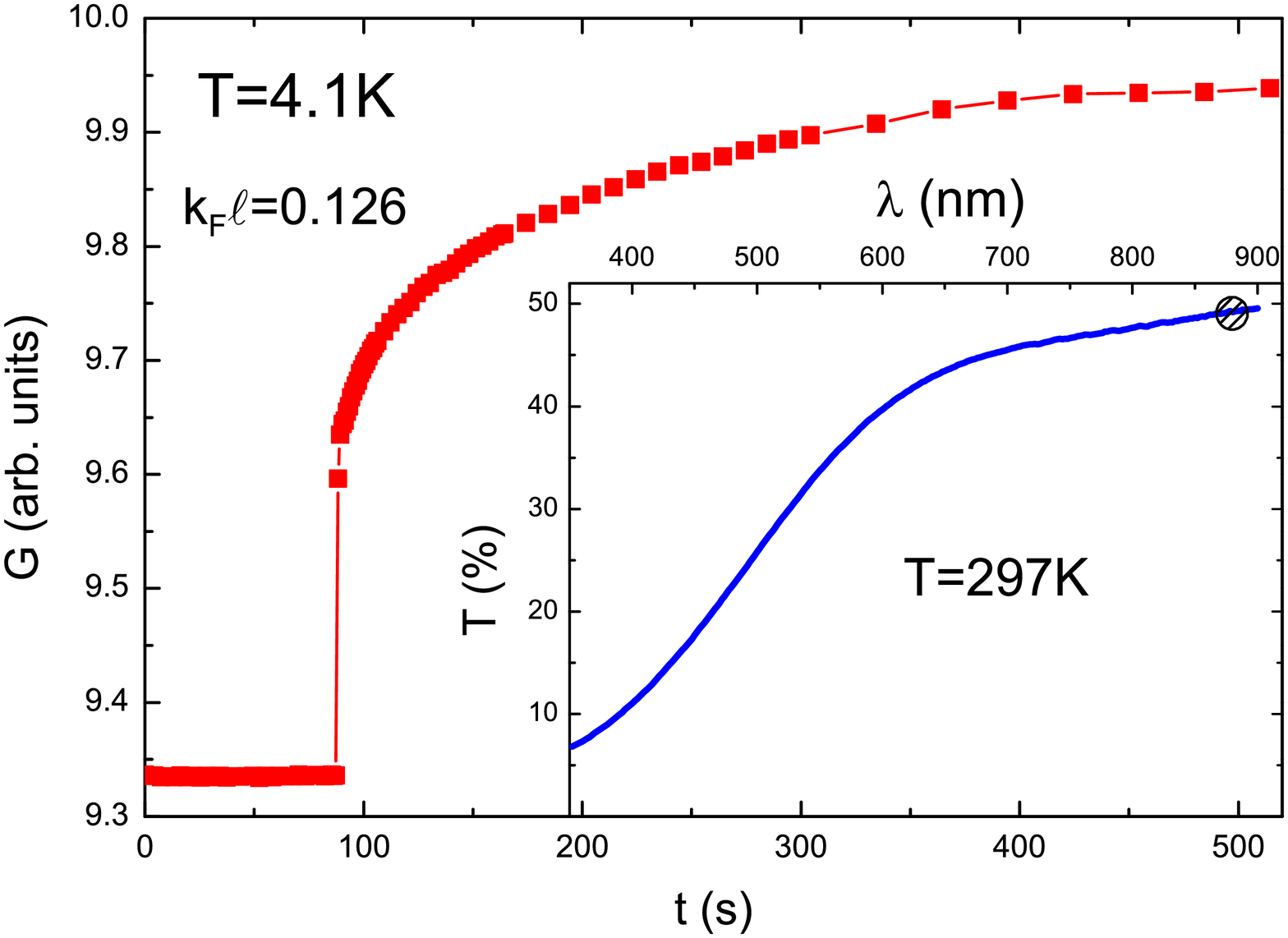}%
\caption{(color online) The temporal dependence of the conductance under the
`extended IR protocol' performed on the same sample (and at the same
conditions except for the duration of exposure) as in Fig.9. The inset shows
the optical transmission through the sample where the spectral part of the
source is delineated by the hatched circle.}%
\end{figure}

The inset In Fig.9 confirms that the relaxation of the excess conductance
created by the IR burst has a logarithmic time dependence with essentially the
same slope as that of the quench-cooling process, and in particular, it
differs from the "persistent-photoconductivity" phenomena well known in
semiconductors \cite{31} in that it apparently decays to its near-equilibrium
value without the need to wait an astronomical time or raise the bath
temperature to regain the pristine, near-equilibrium conductance. However,
only in rare cases is it possible to completely wash-out the MD with just a
brief IR burst. To achieve a complete randomization, required for the
experiment described in the next paragraph, we employ the `extended
IR-protocol' shown in Fig.10. Here the IR radiation is kept for few minutes
while G is monitored and is observed to monotonously increase. The situation
is qualitatively similar to that encountered with either, the
`stress-protocol' or when the bath temperature is raised by a small $\Delta$T,
in both cases the excitation involves generation of phonons \cite{30}. The
IR-protocol is very efficient; it destroys the MD in a matter of minutes under
a moderate change of G in the excitation stage, which is not possible with any
of the "thermal" methods. Using the IR for erasing the MD allows us to learn
about the spatial range of the disturbance created by changing the gate
voltage with a much better time resolution than possible with the
quench-cooling protocol (where one has to wait longer for the ambient
temperature to reach its stationary value). The protocol involved in these
experiments is explained next, and the results of this protocol for a
particular sample are shown in Fig.11.%
\begin{figure}[ptb]%
\centering
\includegraphics[
height=2.559in,
width=3.4255in
]%
{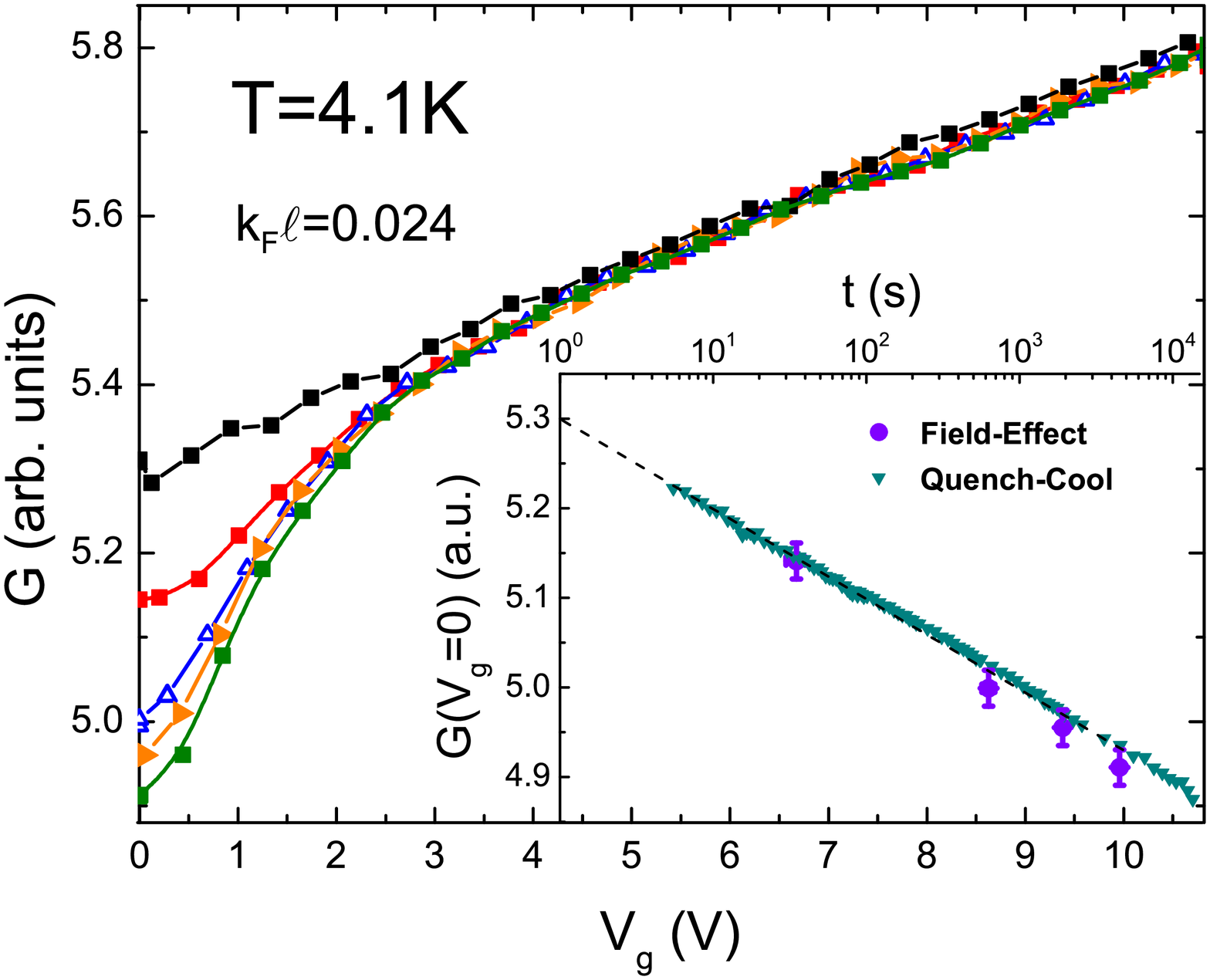}%
\caption{(color online) The combined IR-gate protocol (see text for details)
performed on the sample with the highest disorder in the series. The inset
shows the logarithmic time-relaxation of the V$_{\text{g}}$=0 state taken from
the G(V$_{\text{g}}$) traces as well as the G(t) relaxation from a
quench-cooling protocol (from $\approx$90K to 4.1K). Measurements are with
frequency of 4.8Hz. Note the absence of time dependence in G(V$_{\text{g}}$)
at large V$_{\text{g}}$.}%
\end{figure}

Starting from a near-equilibrium state with the sample under V$_{\text{g}}$=0,
the IR source is turned on for $\approx$10 minutes. Then, V$_{\text{g}}$ is
quickly swept to V$_{\text{g}}$=10.8V while G(V$_{\text{g}}$) is recorded. The
gate voltage is now quickly swept back and parked at V$_{\text{g}}$=0 for a
certain time, allowing the sample to resume the build-up of a memory-dip at
this gate voltage. Next, another V$_{\text{g}}$ sweep is taken and a new
G(V$_{\text{g}}$) is obtained, now exposing a "half-MD". (In these experiments
we measure just the positive V$_{\text{g}}$ range of the G(V$_{\text{g}}$)
trace for the sake of minimizing the time the system is disturbed by the
"exploratory" sweeps). The process is then repeated at different time intervals.

Note first that the conductance at V$_{\text{g}}$=0 decreases logarithmically
for both G(t,V$_{\text{g}}$=0) taken by the field-effect sweeps and G(t) for
the quench-cooling protocol. Moreover, both relaxations proceed at essentially
the same rate (in terms of $\Delta$G/decade). This demonstrates that the
extended IR-protocol could be as efficient as the cooling-protocol in
randomizing the entire volume of the system. The IR protocol is however much
more convenient, quick and less susceptible to cause sample-damage than the
thermal recycling involved in the cooling-protocol.

Turning now to the details of the main plot it is of interest to find out what
is the range in G(V$_{\text{g}}$) over which the modulation due to the MD is
appreciable. Adopting the scheme proposed in \cite{26}, we take this range to
be defined by the gate-voltage V* where the G(t,V$_{\text{g}}$) plots merge
with the `thermodynamic' curve. The signal-to-noise ratio in the data of
Fig.10 (measured at low frequency because of the relatively large R of this
particular sample), makes it difficult to put a value for V*. The IR+gate
protocol was therefore repeated on samples with smaller resistivities and the
results of the combined IR+gate protocol for two of them are shown in Fig.12.%
\begin{figure}[ptb]%
\centering
\includegraphics[
height=2.559in,
width=3.4255in
]%
{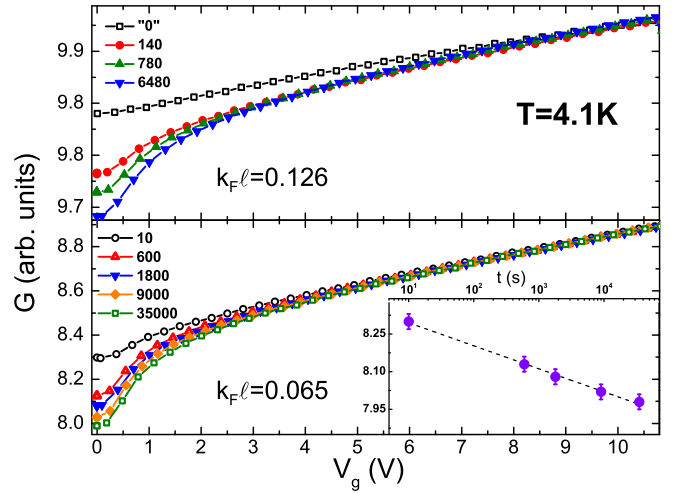}%
\caption{(color online) The combined extended-IR-gate protocol (see text for
details) performed on two samples with reduced disorder and measured at 36Hz
and 74Hz for the sample with k$_{\text{F}}\ell$=0.065 and k$_{\text{F}}\ell
$=0.126 respectively. The inset for the k$_{\text{F}}\ell$=0.065 illustrates
the logarithmic dependence of G(t,V$_{\text{g}}$=0). Note the absence of time
dependence in G(V$_{\text{g}}$) at large V$_{\text{g}}$.}%
\end{figure}

There are two feature revealed by these data. The first is that the modulation
associated with the MD extends to V$_{\text{g}}$ of the order of 6-7~volts,
which is about an order of magnitude higher than estimated from the
`half-height' width that was the basis of the energy-scale estimate of Vaknin
et al \cite{25}. This extended range may also be recognized from the data in
Fig.4 above. As pointed out in \cite{26} this result makes the range in energy
associated with the MD in line with the expectation of models based on the
Coulomb-gap being the mechanism underlying the memory-dip \cite{19,20,21}.

The second feature is more intriguing: The G(t,V$_{\text{g}}$) traces in
Fig.11 and Fig.12 converge to common value for V$_{\text{g}}\geq$V*,
independent of the time that elapsed since the system began relaxing after
being randomized by either the quench-cooling or the IR-protocol. During this
time, the system was relaxing at V$_{\text{g}}$=0 except for the brief
sojourns made by sweeping V$_{\text{g}}$ towards 11-12V, that interrupted the
relaxation by inserting new charges into the sample. The time independence of
the G(t,V$_{\text{g}}$%
$>$%
V*) traces therefore suggests that the gate sweep affects the \textit{entire}
volume of the sample.

This is not a trivial result: It might have been expected that, since the time
of sweep is shorter than the "Maxwell-time", the excess charge induced by
$\delta$V$_{\text{g}}$ resides in the layer of extent $\lambda$, the
Thomas-Fermi length (which is $\approx$10\AA \ in this system), at the
sample-spacer interface, and the parts of the sample that are outside of the
Coulomb interaction range cannot be aware of the disturbance caused by
$\delta$V$_{\text{g}}$. The sample conductance would then continue to relax
uninterrupted by the gate. This will then result in G(t) traces that for
V$_{\text{g}}$%
$>$%
V* obey a logarithmic relaxation like G(t,V$_{\text{g}}$=0) albeit with a
smaller slope (a behavior seen in granular films thicker than 10nm \cite{32}).
Our results, where the excitation by charge insertion is as complete as
quench-cooling for such thick films, is hard to reconcile with such a simple
Coulomb-induced mechanism by which the system gets out of equilibrium. If the
range of excitation was limited to the spatial range of the Coulomb
interaction, the gate excitation might have been progressively less effective
for film thickness exceeding the hopping length, which is the natural
screening length in VRH.

It is natural to expect that above a certain film thickness the gate-protocol
will be less effective, and studying thicker films would be of interest. It is
not clear however that it is the stationary screening-length that is the
relevant length-scale. This scenario ignores the many-body aspects of the
process of charge insertion into an \textit{interacting} system. Pollak was
the first to recognize the complexity of the situation in his Coulomb-gap
paper \cite{23}. The many-body processes that accompany inserting charge into
the interacting system are complicated; some transitions are induced by direct
Coulomb interactions but polarization fields, may induce further transitions
via dipole forces to lower the system energy \cite{23}. In other words, the
disturbance due to the added charge propagates through the system via both,
Coulomb interaction causing direct transitions, and by `after-shock' events of
various nature moving the disturbance further into the system. A similar
scenario, depicting the spread of the disturbance due to the inserted-charge
as `avalanches', was recently simulated by Palassini and Goethe \cite{33}.

The problem is further compounded by the inhomogeneous nature of the medium,
which is an inherent property of the hopping regime. The widely disparate
values of local conductivities of which the medium is composed introduces a
wide spectrum of times involved in the process of charge spreading. The time
it takes the medium to approach equal potentiality, while apparently \cite{34}
still shorter than the V$_{\text{g}}$ sweep-time, is distributed over a wide
spectrum and cannot be simply related to the Maxwell-time associated with the
measured sample conductivity.

In sum, we presented data on the transport properties of thick In$_{\text{x}}%
$O films. In their insulating phase ( k$_{\text{F}}\ell$%
$<$%
0.3) these films exhibit variable-range-hopping conductivity and
current-voltage characteristics that are commonly found in other insulating
systems at low temperatures. Field effect measurements reveal a well developed
memory-dip which is the identifying feature of all intrinsic electron-glasses
\cite{26}. Following excitation, by either, quench-cooling, exposure to IR
radiation, or a sudden change of carrier-concentration, they exhibit
logarithmic relaxation characteristic of other electron-glasses. Their
non-equilibrium transport features disappear at the metal-insulator
transition. This complements the observation made on 2D systems where the
glass phase is likewise restricted to the strongly localized regime.

Experiments with the combined IR exposure and gate excitation protocol reveal
that inserting charge into the electron-glass is effective in `rejuvenating'
the system over a spatial-range considerably larger than the hopping-length.
This seems to suggest that this is a multi-stage process involving many-body
effects similar to that treated by Levitov and Shytov for a diffusive system
\cite{35}. The localization of the electronic states and the inhomogeneous
nature of the medium make such a treatment a challenging task.

In discussing the reasons for the sluggish relaxation of electron-glasses
\cite{4} it was suggested that building up the Coulomb-gap from the out of
equilibrium, random distribution, involves many more sites than those
participating in dc transport. The long range over which the inserted charge
affects occupation of localized sites found it this work suggests that the
contribution of `dead-wood' regions to the relaxation may be substantial.
Local dynamics in these regions is orders of magnitude slower than that of the
sites that are part of the current-carrying-network. As long as the latter can
communicate, directly or indirectly, with sites in the dead-wood regions, the
time it takes the Coulomb gap to establish itself will be influenced by these
regions and therefore their effect on the relaxation dynamics of the system
has to be considered.

\begin{acknowledgement}
This research has been supported by a grant administered by the Israel Academy
for Sciences and Humanities.
\end{acknowledgement}


\begin{thebibliography}{99}
\bibitem {34}Measurements of the charging current during the gating process
(Fig.4 in \cite{8}) show that, at least most of the charge, gets in faster
that the temporal-resolution of the field-effect experiments.
\end{thebibliography}
\end{document}